\documentclass[prb,twocolumn,showpacs,preprintnumbers,amsmath,amssymb,floatfix,nofootinbib]{revtex4-2}

\usepackage[resetlabels,labeled]{multibib}

\usepackage{graphicx}
\usepackage{dcolumn}
\usepackage{bm}
\usepackage{color}
\usepackage{ulem}
\usepackage{multirow}
\usepackage{gensymb}
\usepackage{epstopdf}
\usepackage[colorlinks=true]{hyperref}
\usepackage[dvipsnames]{xcolor}
\usepackage{amsmath}
\usepackage{amssymb}
\usepackage{textcomp}
\usepackage{placeins}
\usepackage{braket}
\usepackage[dvipsnames]{xcolor}
\usepackage{amsmath}
\usepackage{amssymb}
\usepackage{bm}
\usepackage{epsfig}
\usepackage{graphicx}
\usepackage{color}
\usepackage{hyperref}
\usepackage[toc,page,titletoc]{appendix}
\usepackage[capitalise,nameinlink]{cleveref}

\begin{document}

\def\affiIOFFE{Ioffe Institute, Russian Academy of Sciences, 194021 St.~Petersburg, Russia}

\title{Coherent spin dynamics in ensembles of randomly oriented singly charged colloidal nanoplatelets and nanocrystals.}

\author{Aleksandr~A.~Golovatenko}
\affiliation{\affiIOFFE}
\author{Anna~V.~Rodina}
\affiliation{\affiIOFFE}

\date{\today}

\begin{abstract}

We present a theoretical study of the pump-probe Faraday rotation and ellipticity signals in ensembles of uniaxially anisotropic CdSe nanoplatelets and nanocrystals. We use the Faraday rotation mechanism based on the excitation of negative heavy hole trions for a magnetic field applied in the Voigt geometry. Three types of ensembles with typical spatial distributions of the orientation of the anisotropy axis with respect to the direction of light propagation are considered. Faraday rotation and ellipticity signals are modeled for excitation by single and repeated pump pulses, taking into account the anisotropy of the electron $g$-factor. We show that spin dephasing caused by the electron $g$-factor anisotropy and the arbitrary orientation of nanoplatelets or nanocrystals result only in partial damping of oscillation amplitude in contrast to the dephasing caused by the dispersion of the electron $g$-factor in the ensemble. We demonstrate that regardless of the $g$-factor anisotropy degree the oscillation frequency of the Faraday rotation and ellipticity signals for a randomly oriented ensemble is determined by the transverse electron $g$-factor component. 
\end{abstract}

\maketitle


\section{Introduction}

The first pump-probe studies of the coherent spin dynamics of charge carriers in colloidal CdSe nanocrystals (NCs) were carried out in the pioneering works of the Awschalom group \cite{Gupta1999,Gupta2002}. Further pump-probe studies revealed the influence of electrical \cite{Stern2005} or chemical \cite{Wu2021} charging of CdSe NCs on the spin dynamics of electrons. The difference between coherent spin dynamics in zinc-blende and wurtzite CdSe NCs was studied in Ref. \cite{Zhang2014}. Pump-probe studies of CdSe NCs embedded in a glass matrix were carried out in Ref. \cite{Qiang2022}. Recently, experimental pump-probe techniques tested on CdSe NCs have been used to study the coherent spin dynamics of electrons in quasi-two-dimensional colloidal CdSe nanoplatelets (NPLs) \cite{Feng2020, Meliakov2023}. In Ref.~\cite{Feng2020} it was shown that depending on the thickness of the CdS shell, there could be observed one or two Larmor precession frequencies corresponding to differently confined electrons. In Ref.\cite{Meliakov2023},  coherent spin dynamics of electrons in CdSe NPL of different thicknesses was studied. The anisotropy of the electron $g$-factor in CdSe NPLs was measured by spin-flip Raman scattering Ref.\cite{Meliakov2023}. Comparison of the oscillation frequency in the pump-probe study with the spin-flip Raman scattering data showed that this frequency is close to the Larmor frequency, corresponding to the transverse component of the electron $g$-factor.

The studies mentioned above have raised the following questions. The first question concerns the origin of the nonoscillating component in the Faraday rotation (FR) signal in a transverse magnetic field \cite{Gupta2002,Stern2005,Qiang2022} and of the oscillating component in a longitudinal magnetic field \cite{Gupta2002,Zhang2021}. In Ref.~\cite{Gupta2002} it was tentatively argued that as colloidal NCs do not have a well-defined spatial orientation and the optical selection rules are ill-defined, so that boundaries between Faraday and Voigt geometry are blurred. Other possible origins of the nonoscillating component were discussed later in Ref.~\cite{Stern2005}: hole spins pinned along the anisotropy axis \cite{Crooker1997}, surface carrier trapping \cite{Gupta1999}, or the decay of the nonoscillating exciton population \cite{Tartakovskii2004}.  In Ref.~\cite{Qiang2022} nonoscillating component in the Voigt geometry was ascribed to pinned spins of heavy holes.

Another question concerns the number and values of the observable Larmor precession frequencies in the case of a randomly oriented ensemble of CdSe NCs or NPLs with an anisotropic electron $g$-factor \cite{Stern2005,Chen2004}. In Ref.~\cite{Chen2004} it was argued that the two precession frequencies observed in Ref.~\cite{Gupta2002} correspond to the transverse (perpendicular to the anisotropy axis), $g_{\perp}$, and longitudinal (parallel to the anisotropy axis), $g_{||}$, electron $g$-factor components in anisotropic CdSe NCs. However, later it was shown that these two frequencies correspond to electrons with different spatial confinements within a nanocrystal \cite{Hu2019,Wu2021}. In Ref.~\cite{Stern2005} it was argued that the oscillation frequency of the FR signal should correspond to an intermediate value between $g_{||}$ and $g_{\perp}$ of the ensemble-averaged $g$-factor. In Ref.~\cite{Meliakov2023} we stated that for a randomly oriented ensemble, the oscillation frequency is determined by the transverse electron $g$-factor. In this paper, we present a detailed theoretical treatment that supports this assertion.

In order to answer the questions posed above with confidence, it is necessary first to identify the mechanism responsible for the rotation of the probe pulse polarization plane in colloidal NCs and NPLs.
Resident electrons are known to appear in CdSe NCs and NPLs due to photocharging \cite{feng2017dynamic,Hu2019,Hu2019long}. In the case of CdSe NPLs, the presence of resident electrons can be observed directly through a pronounced trion photoluminescence band at low temperatures \cite{Shornikova2020nl,Antolinez2020}. It is known that resident electrons contribute to the formation of FR and ellipticity signals via the trion mechanism described in Refs.~\cite{Yugova2009,Glazov2012} for epitaxial quantum dots. This mechanism involves the creation of long-lived spin polarization of resident electrons because of strict selection rules for the excitation of negative heavy hole trions (a heavy hole plus two electrons in the singlet state) by a circularly polarized pump pulse.
The approach proposed in Ref.~\cite{Yugova2009} was extended to the case of spherical colloidal NCs with a four-fold degenerate hole (trion) ground state in Ref. \cite{Smirnov2012}. A distinctive feature in this case is the dependence of the sign of the created spin polarization of resident electrons on the area of the pump pulse. 

It should be noted that as long as the hole states remain degenerate, as was considered in Ref.~\cite{Smirnov2012}, different spatial orientations of NCs or NPLs in the ensemble do not play a role.
However, often the degeneracy of the hole ground state is lifted by the NC shape anisotropy or by the presence of a built-in crystal field in the case of NCs with a wurtzite crystal structure \cite{Efros1996}. For example, the splitting of the hole states in wz-CdSe NCs reaches $20-30$ meV \cite{Efros1996}.
Due to the strong shape anisotropy of the CdSe NPLs, the splitting of light- and heavy hole states increases to 200 meV \cite{Ithurria2008, Ithurria2011}. In such anisotropic nanostructures, the ground state of the negative trion is formed by a heavy hole and is doubly degenerate. Due to the selection rules for heavy hole optical transitions, the interaction of the trion with light depends on the spatial orientation of the nanostructure anisotropy $c$-axis. As mentioned above, the electron $g$-factor in CdSe NCs and NPLs is anisotropic $g_{\perp}\neq g_{||}$ \cite{Chen2004, Meliakov2023}. This leads additionally to a dependence of the electron Larmor precession frequency on the spatial orientation of the NPL or NC. Therefore, it is necessary to extend the theoretical models from Refs. \cite{Yugova2009,Smirnov2012} to the case of an arbitrary oriented ensemble of NCs or NPLs with an anisotropic electron $g$-factor to answer the questions posed above.

In this paper, we theoretically study the oscillations of the FR and ellipticity signals in ensembles of colloidal NPLs and NCs with different spatial orientations. In section \ref{model}, we introduce the laboratory and nanostructure coordinate systems, discuss possible orientations of the anisotropy axis in the ensemble, and describe the creation of electron spin polarization by single and repeated pump pulses, its evolution over time, and readout by a probe pulse. In Sec.~\ref{results}, we present the results of the calculation of the initial spin polarization and time dependences of the FR and ellipticity signals. In Sec.~\ref{discussion}, we discuss the obtained results and their possible experimental confirmations.

\section{Theoretical model} \label{model}
 
The main approximations of our theoretical model are the following: we assume that NPLs or NCs are singly charged with a resident electron, and the rotation of the probe pulse polarization plane is controlled by the Larmor precession of resident electrons whose spin polarization is created via excitation of trions \cite{Yugova2009}. This assumption is consistent with the experimental fact that the maximum of the FR signal coincides with the maximum of the trion photoluminescence band in Ref.~\cite{Meliakov2023}. Next, we consider singlet negative trions formed by two electrons and a heavy hole. For these trions, the interaction of carriers with light is determined by a two-dimensional electric dipole lying in a plane perpendicular to the $c$-axis of the NPL or NC. The dipole moment of the trion, directed along the $c$-axis, is zero, i.e. there is no interaction of the trion with the light component polarized along the $c$-axis. We note that the dipole moment along the $c$-axis should be taken into account, for example, in the case of perovskite NCs \cite{Kirstein2023}, where trions are formed by holes from the $s$-type valence band and electrons from the spin-orbit-split conduction subband. 

\begin{figure}[hbt]
	\includegraphics[width=1\columnwidth]{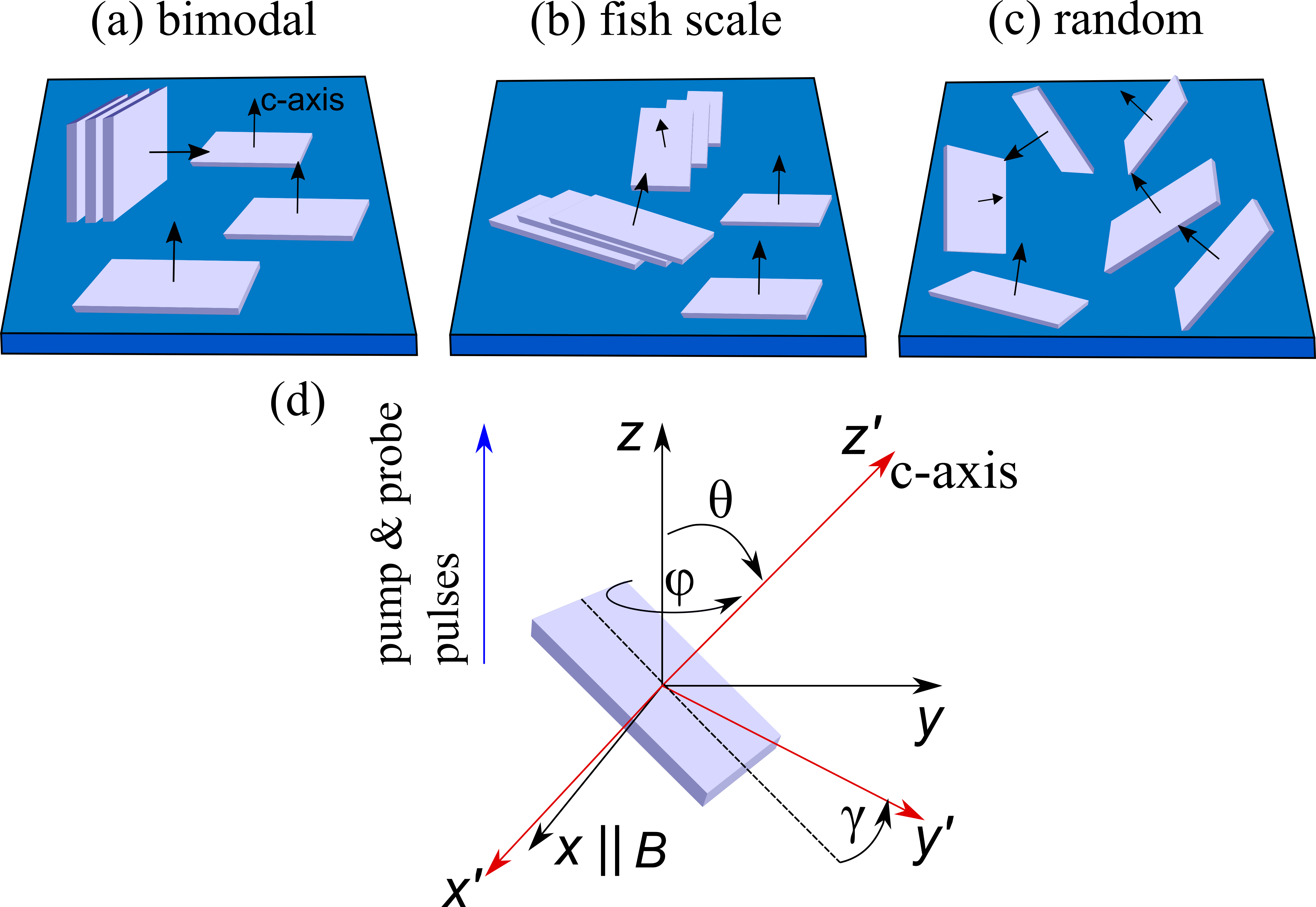}
	\caption{Schematics of nanoplatelet ensembles with different types of spatial orientation: (a)  bimodal, (b)  "fish scale", (c)  random. Panel (d) shows relative orientation of the laboratory and NPL frames. Pump and probe pulses are directed along the $z$-axis. Magnetic field is directed along the $x$-axis. Angles $\varphi, \theta, \gamma$ are nutation, precession and intrinsic angles for the $z-y-z$ Euler rotation sequence.}
	\label{fig:ensembles}
\end{figure}

The lateral size of NPLs is considered to be much smaller as compared to the wavelength of the pump and probe laser pulses. The latter criterion is satisfied for quantum dot-like CdSe NPLs with small lateral sizes used for pump-probe FR studies in Refs.~\cite{Feng2020,Meliakov2023}. For quantum well-like CdSe NPLs \cite{Bouet2013,Vasiliev2015} with lateral dimensions exceeding $100$~nm, this criterion is violated, and another theoretical model should be used. For spherical colloidal NCs, this criterion is always satisfied, since the typical diameter of NCs does not exceed $10$~nm. In the following, we consider NPLs, assuming that the same results are valid for NCs. Possible issues that should be taken into account for NCs are discussed in section \ref{discussion}.
Also, following Refs.~\cite{Yugova2009,Smirnov2012} we neglect local field effects related to the difference between the background dielectric constant of the semiconductor $\varepsilon_{\rm in}$ and the dielectric constant of the matrix $\varepsilon_{\rm out}$.

We assume that the electron spin relaxation time $\tau_{s,e}$ is isotropic and much longer than all other characteristic times of the system, including the laser repetition period $T_R$. In this case, quenching of the FR and ellipticity oscillations is caused by two mechanisms: (i) spin dephasing in the ensemble due to electron $g$-factor anisotropy and different orientations of the anisotropy axis in the ensemble, (ii) spin dephasing due to dispersion of the electron $g$-factor in the ensemble. This assumption also means that the equilibrium spin polarization of electrons along the magnetic field is never reached, and the nonoscillating component of the electron spin polarization along the magnetic field direction remains constant. 

We consider the following NPL ensembles: (i) a bimodal ensemble of NPLs with two preferred orientations (lying flat and standing on their edge), see Fig.~\ref{fig:ensembles}(a);  (ii) an ensemble of NPLs forming a "fish scale" packing, see Fig.~\ref{fig:ensembles}(b). It is assumed that the deviation of the NPL plane in the ensemble from the plane of the substrate is uniformly distributed in the range from 0 to 30 degrees; (iii) an ensemble of NPLs with random orientation of the anisotropy axis, see Fig.~\ref{fig:ensembles}(c). Cases (i) and (ii) are usually realized for NPLs dropcasted on the substrate. Case (iii) can be realized for NPLs in a solvent.

\subsection{Coordinate systems}   

 We start with finding a correspondence between the coordinate system associated with the experimental setup (laboratory frame) and the coordinate system associated with the nanoplatelet (NPL frame). We introduce the laboratory frame axes $X=(x,y,z)$ and the NPL frame axes $X'=(x',y',z')$ shown in Fig.~\ref{fig:ensembles}(d). The direction of the pump and probe pulses coincides with the $z$-axis of the laboratory frame. The laboratory frame and the NPL frame are related by $X^{'}={\textit R}X$, with the following matrix ${\textit R}$:
\begin{widetext}
    \begin{eqnarray}\label{eq:rmatrix}
 {\textit R} = 
\begin{pmatrix}
\cos\varphi\cos\gamma\cos\theta - \sin\varphi \sin\gamma & \quad \cos\gamma \cos \theta \sin \varphi + \cos\varphi \sin\gamma & \quad -\cos \gamma \sin \theta \\
-\cos\gamma \sin \varphi - \cos \varphi \cos \theta \sin \gamma &\quad \cos \varphi \cos \gamma - \cos \theta \sin \varphi \sin \gamma &\quad \sin \theta \sin \gamma\\
\cos \varphi \sin \theta & \quad \sin \varphi \sin \theta & \quad \cos \theta
\end{pmatrix}
  \end{eqnarray}  
 \end{widetext}
The matrix ${\textit R}$ corresponds to the $z-y-z$ rotation sequence that transforms the laboratory frame into the NPL frame. The Euler precession, $\theta$, and nutation, $\varphi$, angles are the polar and azimuthal angles of the NPL anisotropy $c$-axis ($c||z'$) in the laboratory frame, so that the first two rotations entirely define the spatial orientation of the NPL $c$-axis. The third rotation around the $c$-axis by the intrinsic angle $\gamma$ is added for convenience of calculations. It is chosen to satisfy the condition that the $y'$ axis of the NPL frame lies in the $yz$ plane of the laboratory frame. The $y'$ axis can be chosen in this way if the NPLs are axially symmetric with respect to the $c$-axis. Under this assumption, the angle $\gamma$ can be expressed through the angles $\theta$ and $\varphi$ using the condition that the projection of the $x$-axis onto the $y'$-axis is equal to zero, i.e. $R_{2,1}=-\cos\gamma \sin \varphi - \cos \varphi \cos \theta \sin \gamma =0$. We note that the final results of our calculations do not depend on the angle $\gamma$, which confirms the validity of our approach. 

For the chosen coordinate systems, the magnetic field directed along the $x$-axis of the laboratory frame  ${\bm B}_x=B{\bm o_{x}}$ has the following components in the NPL frame:
\begin{eqnarray}
	&&{\bm B}_{x'}=B\sqrt{1-\cos^2\varphi\sin^2\theta}{\bm o_{x'}},\\  \nonumber
	&&{\bm B}_{y'}=0,\\ \nonumber
	&&{\bm B}_{z'}=B\cos\varphi\sin\theta {\bm o_{z'}}.
\end{eqnarray} 
 Here, ${\bm o_{i}}$ and ${\bm o_{i'}}$ ($i=x,y,z$) are unit vectors directed along the coordinate axes of the laboratory and NPL frame, respectively. The value of the $B_{x'}$ component of the magnetic field is determined from the relationship $B_{x'}^2+B_{z'}^2=B^2$.

We also need to find projections on the NPL frame for the circularly polarized pump pulse directed along the $z$-axis of the laboratory frame. We consider further only the $\sigma^+$ polarized pump pulse. This consideration can be easily transferred to the case of $\sigma^-$ polarization.
The $\sigma ^+$ polarized pump pulse in the laboratory frame is described by:
\begin{eqnarray}
    {\bm E}_{\sigma^+}({\bm r},t)=E_p({\bm r},t)\frac{{\bm o_x}+{\rm i}{\bm o_y}}{\sqrt{2}}+c.c.
\end{eqnarray} 

The amplitude $E({\bm r},t)$ is proportional to $e^{-i\omega_p t}$ with $\omega_p$ being the frequency of photons in the pump pulse. In the NPL frame, the pump pulse has both $\sigma^\pm$ components with the following amplitudes:
\begin{eqnarray}
 &&{\bm E}_{\sigma^{+'}}({\bm r},t)=E_p({\bm r},t)\cos^2\frac{\theta}{2}e^{-{\rm i}(\varphi+\gamma)}\frac{{\bm o_{x'}}+{\rm i}{\bm o_{y'}}}{\sqrt{2}}+c.c.,\\
&&{\bm E}_{\sigma^{-'}}({\bm r},t)=-E_p({\bm r},t)\sin^2\frac{\theta}{2}e^{{\rm i}(\gamma-\varphi)}\frac{{\bm o_{x'}}-{\rm i}{\bm o_{y'}}}{\sqrt{2}}+c.c. \nonumber
\end{eqnarray}   
Here, we have neglected the linearly polarized component of the pump pulse ${\bm E}_{z'}({\bm r},t)$ in the NPL frame, since the trion dipole moment along the $z'$-axis is equal to zero.  

Finally, we find components of the probe pulse ${\bm E}_{\rm pr}({\bm r},t)=E_{\rm pr}({\bm r},t){\bm o}_x$ linearly polarized along the $x$-axis of the laboratory frame in the NPL frame. The frequency of the probe pulse $\omega_{\rm pr}$ in general differs from the pump pulse frequency $\omega_{ p}$. Components $E_{x',y',z'}$ can be written similarly to the magnetic field components:
\begin{eqnarray}\label{epump}
	&&{\bm E}_{x'}({\bm r},t)=E_{\rm pr}({\bm r},t)\sqrt{1-\cos^2\varphi\sin^2\theta}{\bm o}_{x'},\\  \nonumber
	&&{\bm E}_{y'}({\bm r},t)=0,\\ \nonumber
	&&{\bm E}_{z'}({\bm r},t)=E_{\rm pr}({\bm r},t)\cos\varphi\sin\theta{\bm o}_{z'}. 
\end{eqnarray} 
Again, the zero dipole moment of the trion along the $z'$ direction  allows us to consider only the $E_{x'}$ component of the electric field. Hence, for the probe pulse, the only important effect arising from the arbitrary orientation of the NPL quantization axis is the renormalization of the component ${\bm E}_{x'}({\bm r},t)$.   

\subsection{Initialization of electron spin polarization}

Following Refs.~\cite{Yugova2009,Smirnov2012} we consider the action of a $\sigma^+$ polarized laser pulse with frequency $\omega_p$ close to the trion resonance frequency $\omega_0$. The pulse duration $\tau_p$ is assumed to be short compared to spin relaxation times of electron and trion, the trion radiative lifetime, and the spin precession period of an electron and a hole in an external magnetic field. 

The pump pulse induces an optical transition between the electron state and the trion state. As a result, a coherent superposition of these states is created, which is described by a four component wavefunction:
\begin{eqnarray}
	\Psi=(\psi_{1/2},\psi_{-1/2}, \psi_{3/2}, \psi_{-3/2}),
\end{eqnarray}
where the $\pm1/2$ subscripts denote the electron spin projection and $\pm3/2$ refer to the spin projection of a hole in a negative trion. The electron spin polarization in the NPL frame is expressed in terms of $\psi_{\pm1/2}$ as:
\begin{eqnarray}
	&&S_{z'}=(|\psi_{1/2}|^2-|\psi_{-1/2}|^2)/2, \\
	&&S_{x'}={\rm Re}(\psi_{1/2}\psi_{-1/2}^*), \nonumber \\ &&S_{y'}=-{\rm Im}(\psi_{1/2}\psi_{-1/2}^*). \nonumber
\end{eqnarray}
 
The change of the $\Psi$ components upon the action of the pump pulse is described by the system of equations: 
\begin{eqnarray}\label{eq:syst}
	&&{\rm i}\dot{\psi}_{3/2}=\omega_0\psi_{3/2}+\cos^2\frac{\theta}{2}f(t)e^{-{\rm i}(\omega_p t-\varphi-\gamma})\psi_{1/2},\\ \nonumber
	&&{\rm i}\dot{\psi}_{1/2}=\cos^2\frac{\theta}{2}f(t)e^{{\rm i}(\omega_p t+\varphi+\gamma})\psi_{3/2},\\ \nonumber
	&&{\rm i}\dot{\psi}_{-3/2}=\omega_0\psi_{-3/2}-\sin^2\frac{\theta}{2}f(t)e^{-{\rm i}(\omega_p t+\varphi-\gamma)}\psi_{-1/2},\\ \nonumber
	&&{\rm i}\dot{\psi}_{-1/2}=-\sin^2\frac{\theta}{2}f(t)e^{{\rm i}(\omega_p t-\varphi+\gamma)}\psi_{-3/2}.
\end{eqnarray} 
Here $\dot{\psi}=\partial \psi/\partial t$, and a smooth envelope of the pump pulse $f(t)$ equals:
\begin{eqnarray}
    f(t)=-\frac{e^{{\rm i}\omega_p t}}{\hbar}\int  d({\bm r})E({\bm r},t)d^3r
\end{eqnarray}

The effective transition dipole $d({\bm r})$ equals \cite{Yugova2009}:
\begin{eqnarray}
   d({\bm r})=-{\rm i}\frac{ep_{cv}}{\omega_0m_0}\int d^3r f_e({\bm r})f_h({\bm r})
\end{eqnarray}
here $p_{cv}$ is the interband matrix element of the momentum
operator taken between the conduction- and valence-band
Bloch functions at the $\Gamma$ point of the Brillouin zone, $f_{e,h}(r)$ are electron and hole envelope wavefunctions. 
From Eqs.~\eqref{eq:syst} one can see that $\sigma^+$ pump pulse creates both trion states $\psi_{\pm 3/2}$ when $\theta \neq 0$. This situation is similar to one considered in Ref.~\cite{Smirnov2012}, where $\sigma^+$ polarized pump pulse excites simultaneously light- and heavy hole trions.

Since the pump pulse interconnects the states $\psi_{1/2}/\psi_{3/2}$ and $\psi_{-1/2}/\psi_{-3/2}$, one can consider these pairs as two independent two-level systems and find the relationship between the electron components of the wavefunction $\Psi$ after,  $\psi_{\pm1/2}(t\rightarrow\infty)$, and before, $\psi_{\pm1/2}(t\rightarrow-\infty)$, the pump pulse. The solution has the following form \cite{Smirnov2012}:
\begin{eqnarray}
	\psi_{\pm1/2}(t\rightarrow\infty)=Q_{\pm}e^{i\Phi_{\pm}}\psi_{\pm1/2}(\rightarrow-\infty).
\end{eqnarray}
Here, the real coefficients $Q_{\pm}$ satisfy the condition $0\leq Q_{\pm}\leq1$ and phases $\Phi_{\pm}$ can be chosen in the range between $-\pi$ and $\pi$. The explicit forms of $Q_{\pm}$ and $\Phi_{\pm}$ depend on the pump pulse shape. In what follows, we consider smooth pulses of the shape proposed by Rosen and Zener \cite{Rosen1932}:
\begin{eqnarray}
	f(t)=\frac{\mu}{\cosh(\pi t/\tau_{p})},
\end{eqnarray}
where the coefficient $\mu$ is a measure of the pulse electric field strength, and $\tau_p$ is the pulse duration. The effective area of the pulse is $\Theta=2\mu\tau_{p}$. For the Rosen-Zener pump pulse parameters $Q_{\pm}$ and $\Phi_{\pm}$ are equal \cite{Smirnov2012}:
\begin{eqnarray}
	Q_{\pm}=\sqrt{1-\frac{\sin^2(\Theta_{\pm}/2)}{\cosh^2\pi \chi}},
\end{eqnarray}  

\begin{eqnarray}
	\Phi_{\pm}={\rm arg}\left[\frac{\Gamma^2(\frac{1}{2}-i\chi)}{\Gamma(\frac{1}{2}-\frac{\Theta_{\pm}}{2\pi}-i\chi)\Gamma(\frac{1}{2}+\frac{\Theta_{\pm}}{2\pi}-i\chi)}\right],
\end{eqnarray}  
here $\Theta_{+}=\Theta\cos^2\theta/2$, $\Theta_{-}=\Theta\sin^2\theta/2$ , and $\chi=(\omega_p-\omega_0)\tau_p/2\pi$ is the dimensionless detuning between the pump pulse and trion frequencies.  

The electron spin polarization components before and after the pump pulse are related by:
\begin{eqnarray}\label{eq:szplus}
	&&S_{z'}^a=\frac{Q_+^2-Q_-^2}{4}+\frac{Q_+^2+Q_-^2}{2}S_{z'}^b, \\
	&&S_{x'}^a=Q_+Q_-(S_{x'}^b\cos(\Phi_+-\Phi_-)+S_{y'}^b\sin(\Phi_+-\Phi_-)), \nonumber \\
	&&S_{y'}^a=Q_+Q_-(S_{y'}^b\cos(\Phi_+-\Phi_-)-S_{x'}^b\cos(\Phi_+-\Phi_-)). \nonumber 
\end{eqnarray}

For $\theta=0$, what is the case of epitaxial quantum dots, $Q_-=1$ and $\Phi_-=0$, and we arrive to Eqs.~16(a-c) from Ref.~\cite{Yugova2009}.

Since we are interested in the ensemble of NPLs, which differ by the spin polarizations $S_{x',y',z'}^b$ before the pump pulse arrival, the averaging over all possible spin polarizations $S_{x',y',z'}^b$ should be done. The averaging results in $S_{x',y',z'}^b=0$. Thus, after a single pump pulse, spin polarization components for a sub-ensemble of NPLs with the same orientation of the $c$-axis are:
\begin{eqnarray}\label{eq:szplusens}
	S_{z'}^a=\frac{Q_+^2-Q_-^2}{4},\quad S_{x'}^a=0,\quad  S_{y'}^a=0. 
\end{eqnarray} 

This result remains true as long as the pump pulse repetition period is much longer than the electron spin relaxation time $\tau_{s,e}$, and when there is no equilibrium spin polarization along the magnetic field direction. 

\subsection{Larmor precession in magnetic field}

In our consideration of the spin precession in the magnetic field we use the conventional assumption \cite{Yugova2009, Smirnov2012} that the hole in the trion looses its spin orientation fast, as compared with the radiative lifetime of the trion. 
In this case, the electron remaining after recombination of a trion is completely depolarized, and the electron spin polarization is determined by Eqs.~ \eqref{eq:szplusens}.
In magnetic field, the electron spin polarization  oscillates according to the Bloch equation:
 \begin{eqnarray}
 	{\bm\dot {\bm S}}+{\bm S}\times{\bm \Omega}+\frac{{\bm S}}{\tau_{s,e}}=0.
 \end{eqnarray}
Here ${\bm S}=(S_{x'},S_{y'},S_{z'})$ and ${\bm \Omega}=(\Omega_{x'},\Omega_{y'},\Omega_{z'})$. $\Omega_{x',y'}=g_{\perp}\mu_BB_{x',y'}/\hbar$ and $\Omega_{z'}=g_{||}\mu_BB_{z'}/\hbar$.

For a magnetic field directed along the $x$ axis of the laboratory frame, and provided that the initial spin polarizations after the pump pulse are $S^a=(0,0, S_{z'}^a)$, the spin precession is described by the equation:
\begin{eqnarray}\label{eq:szonepulse}
	S_{z'}(t)=  S_{z'}^a\left(\frac{\Omega_{z'}^2}{\Omega^2}+\frac{\Omega_{x'}^2}{\Omega^2}\cos\Omega t\right)\exp(-t/\tau_{s,e}).
\end{eqnarray}

One can see that in the case of $B_{z'}\neq0$ there is a nonoscillating component of $S_{z'}(t)\propto\Omega_{z'}^2/\Omega^2$. This result is well known for epitaxial nanostructures in the tilted magnetic field \cite{Marie1999}. The distinctive feature of colloidal NPLs or NCs is that the directions of the magnetic field, pump-probe pulses, and anisotropy axis are mutually noncollinear. For epitaxial nanostructures, as a rule, the anisotropy axis coincides with the propagation direction of pump and probe pulses. 

\begin{figure*}[hbt]
	\includegraphics[width=2\columnwidth]{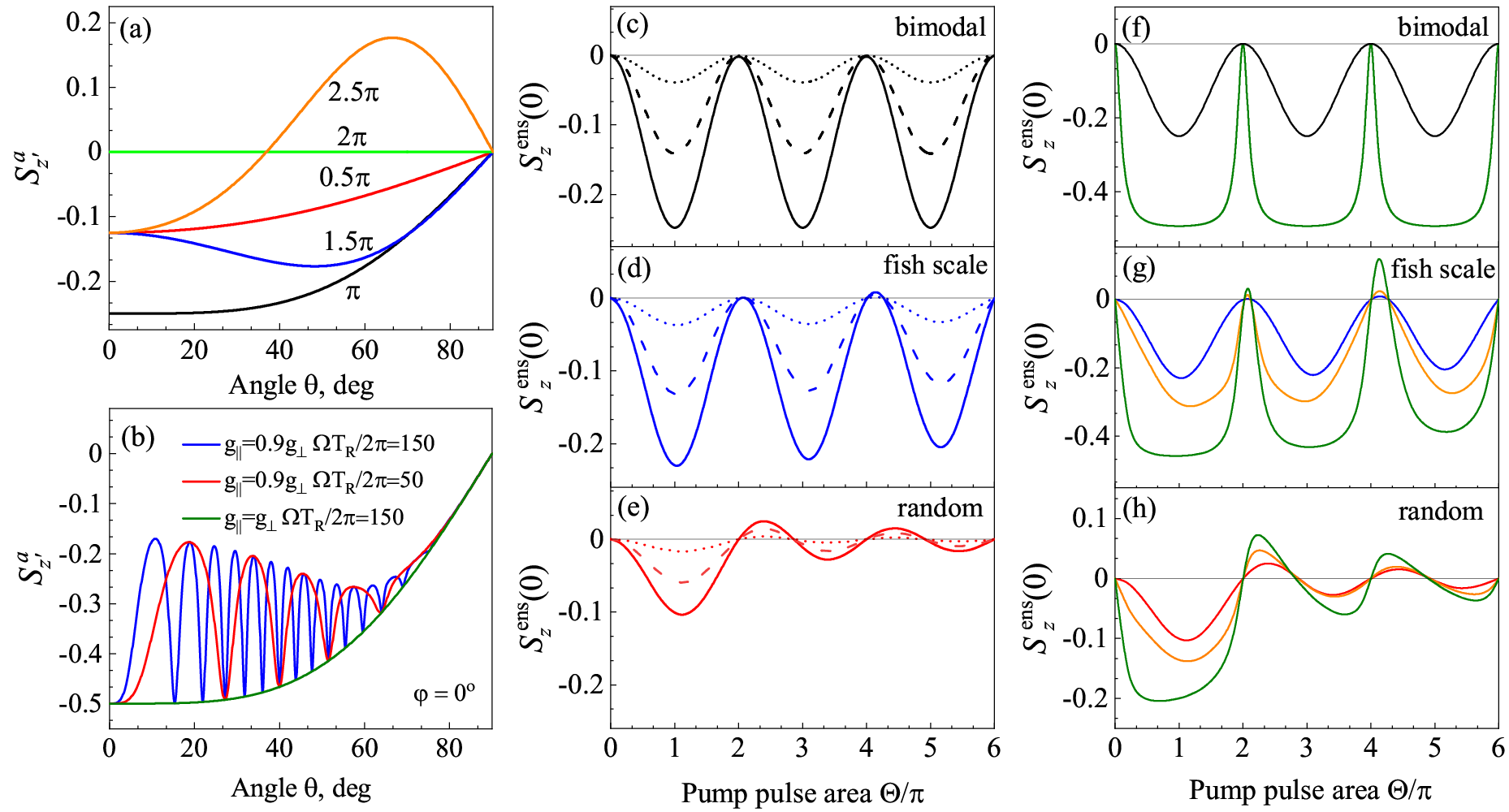}
	\caption{(a) Dependence of the initial spin polarization $S_{z'}^{a}$ after a single pump pulse on the angle $\theta$ for zero detuning $\chi=0$ and different pulse areas $\Theta$. (b) Dependence of the initial spin polarization $S_{z'}^{a}$ after a train of pump pulses on the angle $\theta$ for zero detuning $\chi=0$, pulse area $\Theta=\pi$ and angle $\varphi=0$.  Dependence of the initial ensemble-averaged spin polarization $S_{z}^{\rm ens}(0)$ on the pump pulse area $\Theta$ for the bimodal (c), "fish scale" (d), and randomly oriented ensemble (e). Solid, dashed and dotted lines correspond to the detuning $\chi=0, 0.25, 0.5$, respectively.   Spin polarizations $S_{z}^{\rm ens}(0)$ created by the train of pump pulses with different pulse areas $\Theta$ and $\chi=0$ are shown in panels (f-h). Black, blue and red lines are the same as in panels (c-e) for a single pump pulse. Green and orange lines correspond to ratios of longitudinal and transverse electron $g$-factors $g_{||}/g_{\perp}=1$ and $g_{||}/g_{\perp}=0.9$, respectively. 	 }
	\label{fig:szpth}
\end{figure*}

\subsection{Electron spin polarization created by the train of pump pulses}
If the repetition period of pump pulses, $T_R$, is much faster compared to the electron spin relaxation time $\tau_{s,e}$, spin polarization of electrons can be significantly increased \cite{Kikkawa1998,Glazov2008,Yugova2009}, and the mode-locking of the electron spin coherence can be realized \cite{Greilich2006,Greilich2007}.  This requires that the laser repetition period be equal to an integer number of the electron Larmor precession periods, the so-called phase synchronization condition (PSC). In the case of an arbitrarily oriented NPL, we find the following correspondence between spin polarizations before and after the pump pulse:  
\begin{widetext}
\begin{eqnarray}
    && S_{x'}^b=\frac{\Omega_{x'}^2S_{x'}^a+\Omega_{x'}\Omega_{z'}S_{z'}^a+(\Omega_{z'}^2S_{x'}^a-\Omega_{x'}\Omega_{z'} S_{z'}^a)\cos \Omega T_R-\Omega_{z'}\Omega S_{y'}^a \sin \Omega T_R}{\Omega^2}e^{-T_R/\tau_{s,e}},\\
     && S_{y'}^b=\left( S_{y'}^a\cos\Omega T_R+\frac{(\Omega_{z'} S_{x'}^a-\Omega_{x'}S_{z'}^a)}{\Omega}\sin\Omega T_R\right)e^{-T_R/\tau_{s,e}}, \nonumber \\
     &&S_{z'}^b=\frac{\Omega_{z'} (\Omega_{x'} S_{x'}^a+\Omega_{z'} S_{z'}^a)+\Omega_{x'}(\Omega_{x'} S_{z'}^a-\Omega_{z'} S_{x'}^a)\cos \Omega T_R+\Omega_{x'}\Omega S_{y'}^a \sin \Omega T_R}{\Omega^2}e^{-T_R/\tau_{s,e}}. \nonumber
\end{eqnarray}
\end{widetext}
Combining these equations with Eqs.~(\ref{eq:szplus}) one can find the dependence of $S_{x',y',z'}^a$ created by the train of pump pulses. Temporal dependence of the $S_{z'}(t)$ given by Eq.~(\ref{eq:szonepulse}) for a single pump pulse is now modified to:
\begin{eqnarray}
    &&S_{z'}(t)=  \bigg(\frac{\Omega_{z'}}{\Omega^2}(\Omega_{x'}S_{x'}^a+\Omega_{z'}S_{z'}^a)+ \frac{\Omega_{x'}}{\Omega}S_{y'}^a\sin \Omega t+ \nonumber \\&&\frac{\Omega_{x'}}{\Omega^2}(\Omega_{x'}S_{z'}^a-\Omega_{z'}S_{x'}^a)\cos \Omega t\bigg)e^{-t/\tau_{s,e}}.
\end{eqnarray}

\begin{figure*}[hbt!]
	\includegraphics[width=2\columnwidth]{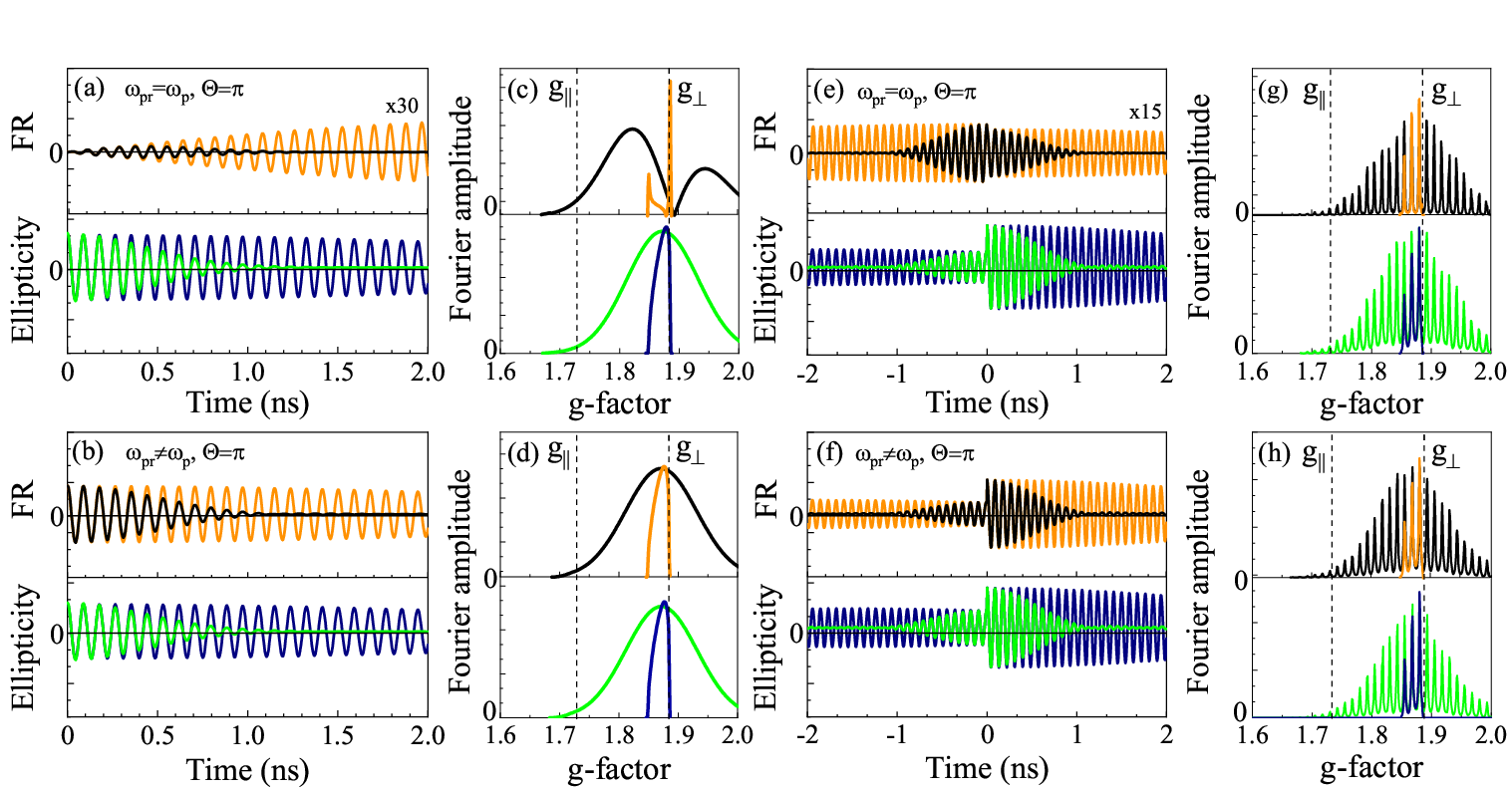}
	\caption{Time traces of the Faraday rotation (black and orange curves) and ellipticity (blue and green curves) signals calculated for the "fish scale" ensemble in the case of a single pump pulse with $\Theta=\pi$ (a,b) and the train of pump pulses (e,f). Panels (c,d,g,h) show the Fourier amplitude spectra of the time traces from panels (a,b,e,f), respectively. Blue and orange time traces are calculated without accounting for the $g$-factor dispersion. Green and black time traces are calculated with $g$-factor dispersion  $\Delta g=0.06$. One color ($\omega_{\rm pr}=\omega_p$) and two color ($\omega_{\rm pr}\neq\omega_p$) pump probe schemes are considered in panels (a,e) and (b,f), respectively.   }
	\label{fig:sztfish}
\end{figure*}

\subsection{Electron spin readout}

The time evolution of the electron spin polarization in an ensemble is detected by the linearly polarized probe pulse. As we have shown above (see Eq.~(\ref{epump})), the tilt of the NPL $c$-axis with respect to the probe pulse direction results in renormalization of the effective electric field in the NPL frame. The renormalized electric field component $E_{x'}({\bm r},t)$ induces dielectric polarizations $\delta P_{x'}$ and $\delta P_{y'}$ along the $x'$ and $y'$ axes of the NPL, respectively~\cite{Yugova2009}. Polarization $\delta P_{x'}$ along the $x'$ axis is proportional to the population of the electron state $n_e=|\psi_{1/2}|^2+|\psi_{-1/2}|^2$. Polarization $\delta P_{y'}$ along the $y'$ axis is proportional to electron spin polarization $S_{z'}(t)$. As the rotation of the probe pulse polarization plane is determined by polarization  along the $y$-axis of the laboratory frame, we need to make a reverse projection of $\delta P_{x',y'}$ onto this axis. 

 Since we consider only the electron contribution to the signal, assuming fast trion recombination ($\approx 100$~ps), the value of $n_{e}$ remains constant. Thus, $\delta P_{x'}$ in each NPL also does not change with time. The contribution of $\delta P_{x'}$ in each NPL to $\delta P_{y}$ is proportional to the product of the $R$ matrix (Eq.~(\ref{eq:rmatrix})) elements $R_{1,1}$ and $R_{1,2}$. Integration over the nutation angle $\varphi$ from $0$ to $2\pi$ results in mutual compensation of contributions from NPLs with different orientations, so that contribution to $\delta P_{y}$ proportional to $n_e$ is absent. However, for an ensemble of NPLs with a nonuniform distribution of angle $\varphi$, a finite nonoscillating value of $\delta P_{y}$ is expected. 

In the case of the $\delta P_{y'}$ polarization component, its contribution to the $\delta P_{y}$ polarization is proportional to the product of $R_{1,1}$ and $R_{2,2}$ matrix elements. Taking into account that $\cos\gamma\sin\varphi=-\cos\varphi\cos\theta\sin\gamma$ we find $\delta P_{y}= \delta P_{y'}\cos\theta$, i.e. it is determined by the projection of the $S_{z'}(t){\bm o}_{z'}$ on the $z$-axis of the laboratory frame. 
The rotation of the probe pulse polarization plane passing through the ensemble of NPLs is controlled by the effective spin polarization of the ensemble:
\begin{eqnarray}
	 &&S_z^{\rm ens}(t)= C\int_{\varphi_1}^{\varphi_2}\int_{\theta_1}^{\theta_2}S_{z'}(t)\cos\theta\sin\theta d\theta d\varphi,
\end{eqnarray}  
where $C=(\varphi_2-\varphi_1)(\cos \theta_1 -\cos \theta_2)$ is the normalization constant.
 
Following \cite{Yugova2009,Smirnov2012}, we calculate the temporal dependence of the FR and ellipticity signals as: 
\begin{eqnarray}
{\cal E} + i {\cal F}=-\frac{3\pi N_{\rm NPL}^{2D}}{q^2 \tau_{rad}}\int \int S_z^{\rm ens}(t)G(\omega_{\rm pr}-\omega_0) \nonumber
\\ \times
\rho_{\Omega}(g_{\perp})\rho_{\rm en}(\omega_{0})dg_{\perp}d\omega_0    
\end{eqnarray}
Here $q=\omega_{\rm pr}\sqrt{\varepsilon_{\rm out}}/c$ is the wave vector of the light in the
matrix, $\tau_{\rm rad}$ is the radiation lifetime of the electro-hole pair, and $N_{\rm NPL}^{2D}$ is the two-dimensional density of NPLs.
The function $G(\omega_{\rm pr}-\omega_0)$ describes the spectral sensitivity of the FR and ellipticity signals.
Explicit forms of $G(\omega_{\rm pr}-\omega_0)$ for different shapes of probe pulses, including the Rosen-Zener pulse, are given in Ref.~\cite{Yugova2009}. Functions $\rho_{\Omega}(g_{\perp})$ and $\rho_{\rm en}(\omega_{0})$ describe the spread of electron $g$-factors and trion resonance frequencies in the ensemble: 
 \begin{eqnarray}
    &&\rho_{\Omega}(g_{\perp})=\frac{1}{\sqrt{2\pi}\Delta g_{\perp}}\exp\left(-\frac{(g_{\perp}-\overline{g}_{\perp})^2}{2(\Delta g_{\perp})^2}\right), \\
     &&\rho_{\rm en}(\omega_{0})=\frac{1}{\sqrt{2\pi}\Delta \omega_0}\exp\left(-\frac{(\omega_0-\overline{\omega}_{0})^2}{2(\Delta \omega)^2}\right)
\end{eqnarray}

For calculation of the time traces of the FR and ellipticity signals, we use a magnetic field $B=0.43$~T, the trion energy in 3ML thick NPLs $\overline{\omega}_0\hbar=2.755$~eV and $\Delta \omega\hbar=10$~meV from Ref.~\cite{Meliakov2023}. 
For the transverse component of electron $g$-factor we use its dependence on the trion energy from Ref.~\cite{Meliakov2023}:
\begin{eqnarray}
    \overline g_{\perp}(\hbar \omega_0)=A\hbar \omega_0+C
\end{eqnarray}
with $A=0.6$~eV$^{-1}$ and $C=0.232$.  We also use the fixed ratio of the $g$-factor components $g_{||}/g_{\perp}=0.92$ \cite{Meliakov2023} and assume that it is constant in the ensemble. The standard deviation for the $g$-factor distribution $\Delta g_{\perp}=0.06$ determined in Ref.~\cite{Meliakov2023} is used.For calculations with a train of pump pulses, a laser repetition period of $T_R=13.2$~ns is used, which corresponds to a repetition frequency of $76$~MHz.

\section{Results}\label{results}

\begin{figure*}[hbt]
	\includegraphics[width=2\columnwidth]{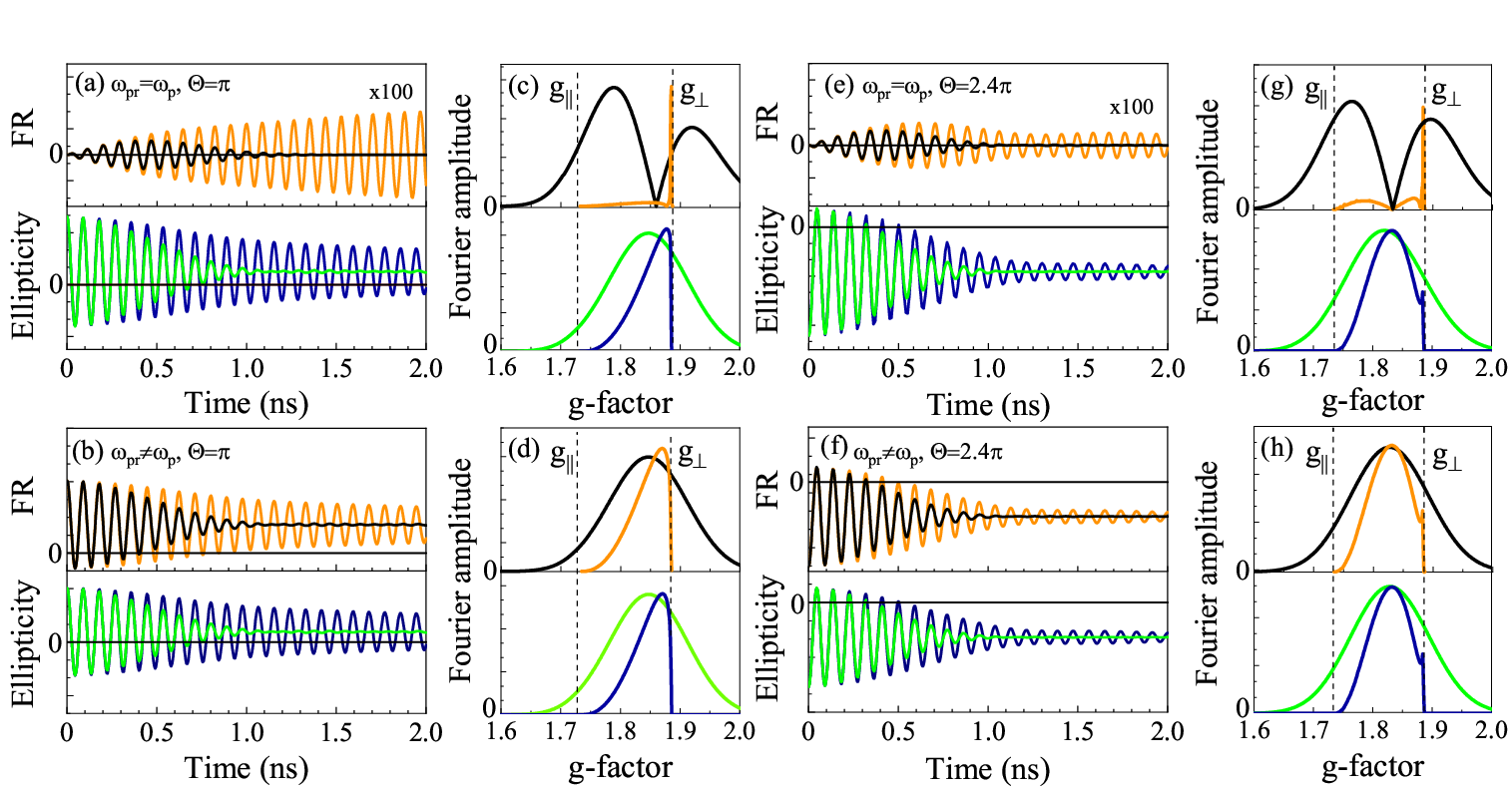}
	\caption{Time traces of the Faraday rotation (black and orange curves) and ellipticity (blue and green curves) signals (a,b,e,f) and corresponding Fourier amplitudes spectra (c,d,g,h) for the randomly oriented ensemble in the case of single pump pulse. Blue and orange time traces are calculated without accounting for the $g$-factor dispersion. Green and black time traces are calculated with $g$-factor dispersion  $\Delta g=0.06$. The pump pulse area $\Theta$ equals to $\pi$ and $2.4\pi$ in panels (a,b) and (e,f), respectively. One color ($\omega_{\rm pr}=\omega_p$) and two color ($\omega_{\rm pr}\neq\omega_p$) pump probe schemes are considered in panels (a,e) and (b,f), respectively.  }
	\label{fig:sztrand1}
\end{figure*}

At first, we calculate the dependence of the initial spin polarization $S_{z'}^a$ on the angle $\theta$ for different effective pump pulse areas $\Theta$ and zero detuning $\chi$. The calculated dependences are shown in Figure \ref{fig:szpth}(a). One can see that for $\theta<\pi/6$, spin polarization created by the $\pi$-pulse remains almost the same. For larger $\theta$, spin polarization decreases and becomes zero at $\theta=\pi/2$. The $2\pi$-pulse results in $S_{z'}^a=0$ for all angles $\theta$ (see the green curve in Fig.~\ref{fig:szpth}(a)).  A further increase of $\Theta$ results in the formation of two regions with opposite signs of $S_{z'}^a$: negative spin polarization in NPLs with a small tilt angle $\theta$, and positive spin polarization in NPLs with $\theta>\pi/6$ (see orange curve in Fig.~\ref{fig:szpth}(a)). 

The dependence of the initial spin polarization on the angle $\theta$ in the case of repeated pump pulses is shown in Fig.~\ref{fig:szpth}(b) for the angle $\varphi=0$ and the pump pulse area $\Theta=\pi$. When the PSC are fulfilled, the angular dependence for an isotropic $g$-factor (green curve) reminds the amplified dependence for a single pump pulse. Anisotropy of the electron $g$-factor results in amplification of $S_{z'}^a$ in a discrete number of angles $\theta$ satisfying the PSC (red and blue curves). 

\begin{figure*}[hbt!]
	\includegraphics[width=2\columnwidth]{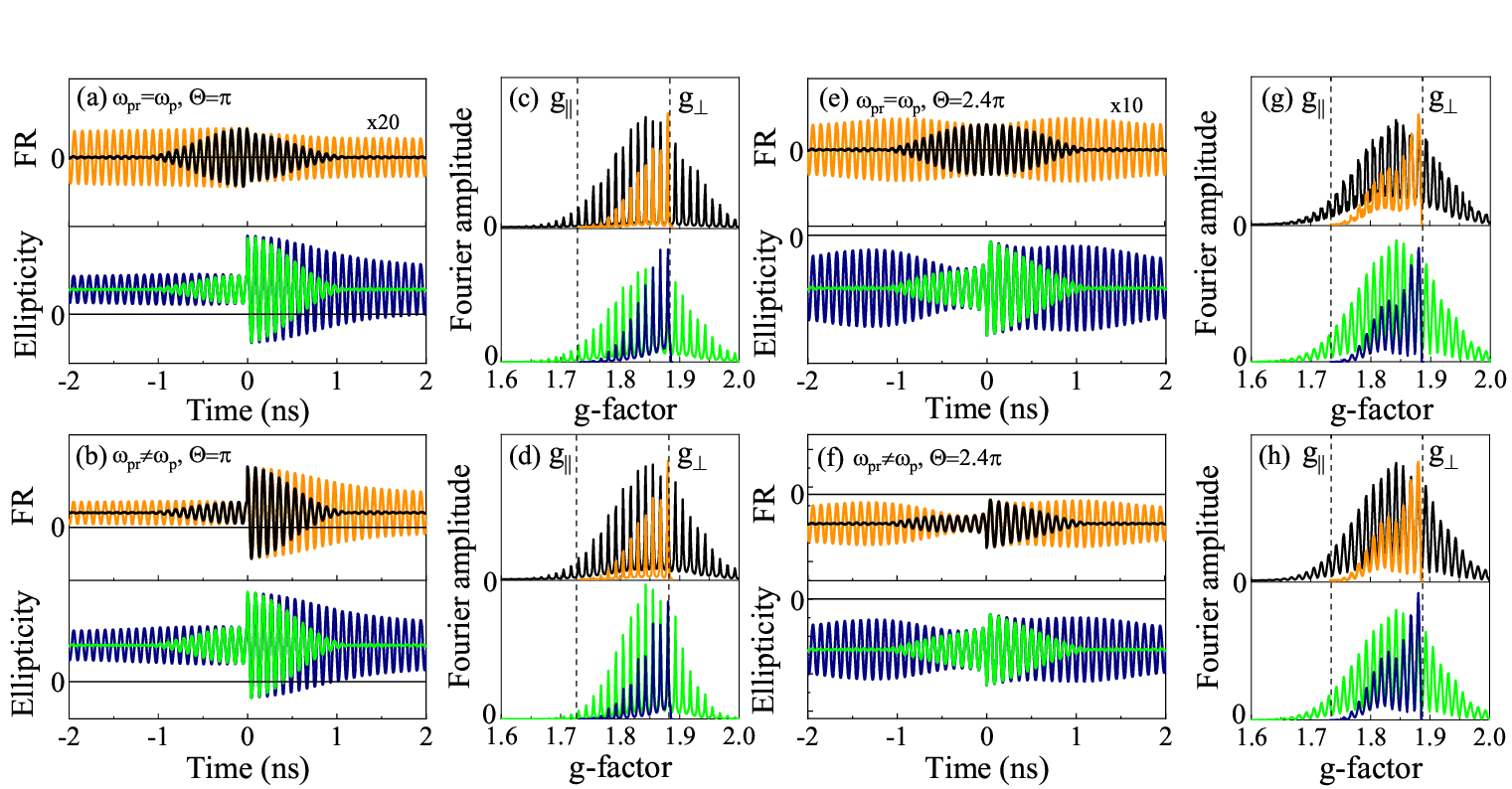}
	\caption{Time traces of the Faraday rotation (black and orange curves) and ellipticity (blue and green curves) signals (a,b,e,f) and corresponding Fourier amplitude spectra (c,d,g,h) for the randomly oriented ensemble in the case of the train of pump pulses. Blue and orange time traces are calculated without accounting for the $g$-factor dispersion. Green and black time traces are calculated with $g$-factor dispersion  $\Delta g=0.06$. The pump pulse area $\Theta$ equals to $\pi$ and $2.4\pi$ in panels (a,b) and (e,f), respectively. One color ($\omega_{\rm pr}=\omega_p$) and two color ($\omega_{\rm pr}\neq\omega_p$) pump probe schemes are considered in panels (a,e) and (b,f), respectively.}
	\label{fig:sztrand2}
\end{figure*}

The ensemble-averaged spin polarization $S_z^{\rm ens}(0)$ created by a single pump pulse calculated for different pulse areas in the three types of NPL ensembles are shown in Fig.~\ref{fig:szpth}(c-e). For the bimodal distribution, the dependence is the same as it were an ensemble of epitaxial QDs ~\cite{Yugova2009}. This is due to the zero spin polarization $S_{z'}^a$ created by the pump pulse in vertically standing NPLs. On the other hand, the subensemble of NPLs lying flat on the substrate is identical to an ensemble of epitaxial QDs. 
The bimodal ensemble of NPLs shows typical Rabi oscillations of the spin polarization created by the pump pulse (see Fig.~\ref{fig:szpth}(c)). 
Spin polarization $S_z^{\rm ens}(0)$ in the ensemble of NPLs with the "fish scale" packing is shown in Fig. \ref{fig:ensembles}(d). 
The dependence of the $S_{z}^{\rm ens}(0)$ on the pump pulse area is similar to that for flat-lying NPLs, except that for large pump pulse areas there are regions with opposite sign of the electron spin polarization.
For the randomly oriented ensemble, the most striking feature is a periodic sign-reversal dependence of the initial spin polarization on the pump pulse area (see Fig.~\ref{fig:szpth}(e)). This sign-reversal behavior is similar to that obtained in Ref.~\cite{Smirnov2012} for spherical NCs with degenerate states of light- and heavy hole trions.
An increase in the detuning $\chi$ results in a gradual decrease of the created spin polarization in all three types of ensembles (see dashed and dotted curves in Fig. ~\ref{fig:szpth}(c-e)).

The initial spin polarizations $S_z^{\rm ens}(0)$ created by the train of pulses are shown in Fig.~\ref{fig:szpth}(f-h). 
The rise in the initial spin polarization for the bimodal ensemble under the PSC is shown in Fig.~\ref{fig:szpth}(f). For the "fish scale" and random ensembles, we find that an increase in the initial spin polarization depends on the ratio between longitudinal and transverse $g$-factors of the electron. The largest $S_z^{\rm ens}(0)$ is achieved in the case $g_{||}=g_{\perp}$ when for all electrons in the ensemble the PSC is fulfilled (green curves in Fig.~\ref{fig:szpth}(g,h)). Anisotropy of the electron $g$-factor results in different Larmor precession frequencies in ensemble and consequently decreases the gain in $S_z^{\rm ens}(0)$ achieved by repeated pump pulses (orange curves in Fig.~\ref{fig:szpth}(g,h)).

As the bimodal ensemble is equivalent to an ensemble of epitaxial QDs studied in Ref. \cite{Yugova2009}, we exclude it from further analysis of the FR and ellipticity time traces and focus on the ensembles with the "fish scale" and the random orientations.
The time evolution of the FR and ellipticity signals for the "fish scale" ensemble is shown in Fig.~\ref{fig:sztfish} for the pump pulse area $\Theta=\pi$. As shown in Fig.~\ref{fig:sztfish}(a), for a single pump pulse with $\hbar\omega_{p }=\hbar\omega_{\rm pr}=\hbar\overline{\omega}_0=2755$~meV the FR signal (orange curve) is negligibly small, compared to the ellipticity signal. It is a consequence of the asymmetric behavior of the imaginary part of the $G(\omega_{\rm pr}-\omega_0)$ function, as was shown in Refs.~\cite{Yugova2009, Smirnov2012}. The FR signal is precisely zero at $t=0$ and gradually increases due to the spectral dependence $\overline g_{\perp}(\hbar \omega_0)$. 
In the case of the ellipticity signal (blue curve in Fig.~\ref{fig:sztfish}(a)) the slow decrease in the oscillation amplitude is caused by the anisotropy of the $g$-factor and the different spatial orientation of the NPLs in the ensemble.
When the dispersion of the electron $g$-factor is included, the corresponding spin dephasing mechanism results in full quenching of the FR and ellipticity signals already at $t=1$~ns (see green and black curves in Fig.~\ref{fig:sztfish}(a)).

The presence of a small nonoscillating component in the ellipticity signal can be observed in Fig.~\ref{fig:sztfish}(a). This component is caused by NPLs with the anisotropy axis noncollinear to the magnetic field direction. This nonocsillating component will occur even in an ensemble of electrons with an isotropic electron $g$-factor. 
In contrast to the ellipticity signal, the nonoscillating component is absent in the FR signal, again, due to the asymmetric shape of the $G(\omega_{\rm pr}-\omega_0)$ function. 

For frequency domain analysis of the FR and ellipticity signals, we calculate semi-analytically corresponding Fourier amplitude spectra, $A_F(g)$, by summation of spectra from individual NPLs with all possible orientations and resonant trion frequencies $\omega_0$. We calculate $A_F(g)$ as a function of the effective $g$-factor instead of the frequency $\Omega$ using the relationship $g=\Omega\hbar/\mu_BB$. The Fourier spectrum of an ensemble can be easily calculated, since the Fourier image of the cosine function in the time domain (see Eq.~\ref{eq:szonepulse}) is the delta function in the frequency domain. The $A_F(g)$ spectra of the FR and ellipticity signals from Fig.~\ref{fig:sztfish}(a) are shown in Fig.~\ref{fig:sztfish}(c).  The Fourier spectrum of the FR signal in the absence of the dispersion of the electron $g$-factor consists of two sharp peaks (see the orange curve in Fig.~\ref{fig:sztfish}(c)) with opposite phases. Taking into account the dispersion of the $g$-factor leads to the formation of two broad peaks in the Fourier spectrum (black curve in Fig.~\ref{fig:sztfish}(c)).  We find that in the absence of the $g$-factor dispersion the Fourier spectrum of the ellipticity signal is asymmetric with a peak close to the transverse electron $g$-factor (blue curve in Fig.~\ref{fig:sztfish}(c)). The account of the dispersion of the electron $g$-factor results in a broadening of the $A_F(g)$ spectrum, but its maximum remains near $g_{\perp}$.

When the probe pulse frequency $\omega_{\rm pr}$ is detuned from the pump pulse frequency $\omega_p$ ($\hbar \omega_{\rm pr}=2756.1$~meV), the FR and ellipticity time traces and the corresponding $A_F(g)$ spectra become similar (see Fig.~\ref{fig:sztfish}(b,d)). 

In Fig.~\ref{fig:sztfish}(e,f) time traces of the FR and ellipticity signals in the case of a train of pump pulses are shown. One can see that oscillation amplitudes between pump pulses remain finite when only dephasing due to the $g$-factor anisotropy and different spatial orientations of NPLs is considered. When the $g$-factor dispersion is included, we find the full quenching of the oscillations after a previous pump pulse and their revival before the next pump pulse (black and green curves in Fig.~\ref{fig:sztfish}(e,f)). The $A_F(g)$ spectra consist of a discrete number of peaks (see Figs.~\ref{fig:szpth}(g,h)) centered at $g$-factor values satisfying the PSC. 

Time traces of the FR and ellipticity signals for the randomly oriented ensemble are shown in Fig.~\ref{fig:sztrand1}. For the pump pulse area $\Theta=\pi$ time traces calculated with $\omega_{\rm pr}=\omega_p$ and $\omega_{\rm pr}\neq\omega_p$, are shown in Fig.~\ref{fig:sztrand1}(a,b), respectively. Compared to the "fish scale" ensemble, the spin dephasing caused by the $g$-factor anisotropy is accelerated due to a larger spread of Larmor frequencies in the randomly oriented ensemble. The relative amplitude of the nonoscillating signal component is also increased. For the FR signal calculated with $\omega_{\rm pr}=\omega_p$ we find the same behavior as for the "fish scale" ensemble: the nonoscillating component is absent, and the oscillation amplitude is very small, compared to the ellipticity signal. 

The $A_F(g)$ spectra of the time traces from Fig.~\ref{fig:sztrand1}(a,b) are shown in Figs.~\ref{fig:sztrand1}(c,d). The $A_F(g)$ spectrum of the FR signal calculated with $\omega_{\rm pr}=\omega_p$ has a sharp peak at $g_{\perp}$ and a broad low-amplitude shoulder between $g_{||}$ and $g_{\perp}$. We note that the phases of the sharp peak and the broad shoulder are opposite. When the dispersion of the electron $g$-factor is taken into account, the Fourier amplitude spectrum consists of two broad peaks, as in the case of the "fish scale" ensemble (see Fig.~\ref{fig:sztfish}(c)). 
The $A_F(g)$ spectra of the ellipticity signal in Fig.~\ref{fig:sztfish}(c,d) calculated with $\omega_{\rm pr}=\omega_p$, and of the FR and ellipticity signals calculated with $\omega_{\rm pr}\neq\omega_p$ are asymmetric with the maxima near the transverse electron $g$-factor. Consideration of the $g$-factor dispersion results in a shift of the $g$-factor value corresponding to the maximum of the $A_F(g)$ spectra towards $g_||$. This result shows that the effective $g$-factor, which determines the oscillation frequency of the ensemble, differs from the transverse $g$-factor of the electron by the amount of the $g$-factor dispersion.

As shown in Fig.~\ref{fig:szpth}
(e), in the randomly oriented ensemble initial spin polarization $S_z^{\rm ens}(0)$ with opposite sign can be created when $\Theta=2.4\pi$.
The time traces of the FR and ellipticity signals for the pump pulse area $\Theta=2.4\pi$ are shown in Fig.~\ref{fig:sztrand1}(e,f) for $\omega_{\rm pr}=\omega_{p}$ and for $\omega_{\rm pr}\neq\omega_{p}$, respectively. It can be seen that the nonoscillating components of the ellipticity signal in Fig.~\ref{fig:sztrand1}(e), and of the FR and ellipticity signals in Fig.~\ref{fig:sztrand1}(f) have opposite signs, compared to the signals excited by the pump pulse with $\Theta=\pi$. The $A_F(g)$ spectra of these signals in contrast to the case with excitation by the pump pulse with $\Theta=\pi$ have peak positions at $g\approx1.83$, i.e. in between $g_{||}$ and $g_{\perp}$. Taking into account the $g$-factor dispersion broadens the Fourier spectra, but does not shift their peak positions.
The FR signal calculated with $\omega_{\rm pr}=\omega_{p}$ (orange curve in Fig.~\ref{fig:sztrand1}(e)) has zero non-oscillatory component and shows periodic beats. These beats can be understood by looking at the Fourier spectrum shown by the orange curve in Fig.~\ref{fig:sztrand1}(g), which consists of two relatively broad peaks and one sharp peak centered at different values of the $g$-factor.

The oscillations of the FR and ellipticity signals in the case of the train of pump pulses for the randomly oriented ensemble are shown in Fig.~\ref{fig:sztrand2}. Compared to signals excited by a single pump pulse, an increase of the nonoscillating component amplitude is observed, as well as an increase of the FR signal calculated with $\omega_{\rm pr}=\omega_p$ (see Fig.~\ref{fig:sztrand2}(a,e)). 
The $A_F(g)$ spectra of the FR and ellipticity signals with different pump-pulse areas and different relationships between $\omega_{\rm pr}$ and $\omega_{p}$ have similar shapes. The spectra consist of discrete peaks placed at $g$-factor values satisfying the PSC. Without the $g$-factor dispersion a peak with the maximum amplitude is placed at $g_{\rm perp}$. When the dispersion of $g$-factor is included, the $A_F(g)$ spectra are broadened and their maxima are shifted toward $g_{||}$.

\section{Discussion}\label{discussion}

In our calculations, we considered ensembles of NPLs. In the case of NCs ensemble the following remarks should be done. As the shape of NCs is usually close to spherical, one can expect that the random orientation is realized both for NCs in solution and for dropcasted NCs. 
Another point concerning NCs is the large spread of the trion resonance energies $\hbar\omega_0$ ($\approx 50~meV$) for a typical NCs size dispersion $5\%$. The dispersion of resonance frequencies in NCs is comparable to or exceeds the light-heavy trions splitting. As a result, excitation of light-hole trions can give an additional contribution to the FR and ellipticity signals.  
The limiting case of degenerate states of light- and heavy hole trions was considered in Ref.\cite{Smirnov2012}. Our work describes another limiting case with infinite splitting of hole states. It is worth noting that in both limiting cases the dependence of the spin polarization sign on the pump pulse area $\Theta$ is expected. It will manifest itself in a change in the phase of the oscillating component of the signal. In the case of NPL or anisotropic NC we also expect a change in the sign of the amplitude of the nonoscillating component.

We considered the creation of the spin polarization of resident electrons assuming that the spin relaxation of a hole in the trion is fast, while the spin relaxation time of the resident electron is much larger than the trion radiative recombination time. In this case, the created spin polarization corresponds to the spin of the resident electron in nonexcited NPLs or NCs. If the spin relaxation time of a hole is much longer than the trion radiative recombination time, different scenarios are possible. The spin of the electron in the excited structures can still be lost due to a fast nonradiative Auger recombination of the trion, what is the case of the bare-core CdSe NCs \cite{Vaxenburg}. Another possibility is to have the initial spin of the resident electron in the excited NPLs or NCs preserved during the trion lifetime and lost in the nonexcited structures. Such a situation can be realized in zero magnetic field only if the electron spin relaxation time is much smaller than the trion lifetime. If this is not the case, spin polarization of resident electrons can be created only in an applied magnetic field \cite{Glazov2012,Marie1999,Kennedy2006, Dutt2005}. In these NPLs, where the pump pulse does not excite a trion, spin of the resident electron will precess during the trion lifetime. If the transverse $g$-factor of the hole, i.e. of the trion, is close to zero, in excited NPLs or NCs  resident electrons after the trion recombination will have the same spin polarization as before the pump pulse. As a result, spin compensation in the system of resident electrons is disrupted and nonzero spin polarization of resident electrons is created. For the last two cases, the sign of polarization will be opposite as compared to the case of the fast spin relaxation of the hole considered by us.    

In the case of the heavy hole trion in the randomly oriented ensemble, the change in the initial spin polarization sign could be detected from the nonoscillating component which decays with the electron spin relaxation time. The non-oscillating component is caused by the longitudinal component of the magnetic field in the NPL frame with $\theta \neq 0$, as was pointed in Ref.~\cite{Gupta2002}. A nonoscillating FR component was observed for NCs ensembles \cite{Gupta2002,Stern2005,Qiang2022} but was absent for NPLs \cite{Meliakov2023}. This probably indicates that the NPL ensembles studied in \cite{Meliakov2023} have a preferential orientation of the $c$ axis parallel to the direction of the pump and probe pulses (bimodal or "fish scale"). This conclusion agrees with measurements of the electron $g$-factor anisotropy by spin-flip Raman scattering in the tilted magnetic field Ref.~\cite{Meliakov2023}.         

\begin{figure}[hbt!]
	\includegraphics[width=1\columnwidth]{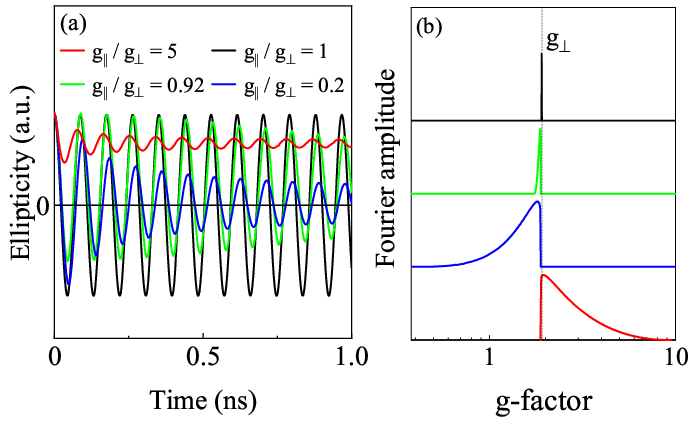}
	\caption{Time traces (a) and Fourier amplitude spectra (b) of the ellipticity signals for the randomly oriented ensembles calculated with fixed $g_{\perp}=1.88$ and different ratios $g_{||}/g_{\perp}$. Excitation by a single pump pulse with $\Theta=\pi$ and $\omega_{\rm pr}=\omega_p$ is considered. }
	\label{fig:hole}
\end{figure}

The above theoretical consideration can be applied to positive trions, containing two heavy holes and one electron.  In the case of heavy holes, large anisotropy of $g$-factor is expected, since the transverse hole $g$-factor differs from zero in the degree of admixture of the light hole states or cubic anisotropy of the hole wavefunction related to the valence band warping or the shape of the confinement potential  \cite{Marie1999,Semina2023}. For clarity, the results of calculation of the ellipticity signal and the corresponding $A_F(g)$ spectra for a fixed $g_{\perp}=1.88$ and various ratios $g_{||}/g_{\perp}$ without accounting for the $g$-factor dispersion are shown in Fig.~\ref{fig:hole}. One can see that time traces have close oscillation frequencies, determined by the shape of the Fourier spectra with maxima at $g_{\perp}$, as shown in Fig.~\ref{fig:hole}(b). 
Thus, for positively charged randomly oriented NPLs or NCs one can expect oscillations of the FR or ellipticity signals with a frequency corresponding to the transverse $g$-factor of the heavy hole, regardless of the $g$-factor anisotropy. Again, the deviation of the effective $g$-factor from $g_{\perp}$ can be caused by the dispersion of the $g$-factor in the ensemble.

Finally, in Fig.~\ref{fig:hole}(a) one can see that the value of the longitudinal component $g_{||}$ affects the amplitude of the nonoscillating component and the dephasing rate. For $g_{\perp}\gg g_{||}$ and $g_{\perp}\approx g_{||}$ the amplitude of the oscillating component is small. When $g_{\perp}\ll g_{||}$, as expected for heavy holes, the nonoscillating component dominates. Since the nonoscillating component of the signal in Refs.~\cite{Gupta2002,Qiang2022} is relatively small, and the oscillations occur at the Larmor frequency of the electron, it is most natural to conclude that the nonoscillating component is also due to electrons rather than holes.    

\section{Conclusion}

The present study shows that coherent electron spin dynamics in ensembles of randomly oriented, singly charged colloidal nanoplatelets and nanocrystals can exhibit additional features that are absent in epitaxial quantum dots, namely: (i) presence of the nonoscillating component of the Faraday rotation or ellipticity signal which amplitude changes sign with an increase in the pump pulse area, (ii) dependence of the frequency and the phase of the oscillating signal component on the pump pulse area, (iii) oscillation frequency is close to the transverse electron $g$-factor for any degree of the $g$-factor anisotropy in an ensemble without $g$-factor dispersion excited by the $\pi$-pulse, (iv) presence of an additional spin dephasing mechanism contributing to the $T_2^*$ spin relaxation time  via  anisotropy of the electron  $g$-factor and different spatial orientations of nanoplatelets or nanocrystals in ensemble.

\section{Acknowledgments}
We thank M.M. Glazov, D.S. Smirnov, E.A. Zhukov, I.A. Yugova and D.R. Yakovlev for valuable discussions. The work was supported by the Russian Science Foundation (Grant No.
23-12-00300).

\end{document}